\documentclass[prc,twocolumn,superscriptaddress,showpacs,amssymb,amsmath,amsfonts,aps]{revtex4}
\usepackage{graphicx}
\usepackage{dcolumn}
\usepackage{epsfig}
\begin{document}
\title{Electroexcitation of nucleon resonances of the $[70,1^-]$ multiplet
in a light-front relativistic quark model \\}

\newcommand*{\JLAB }{ Thomas Jefferson National Accelerator Facility,
Newport News, Virginia 23606, USA}
\affiliation{\JLAB }
\newcommand*{\YEREVAN }{ Yerevan Physics Institute, 375036 Yerevan,
Armenia}
\affiliation{\YEREVAN }
\author{I.G.~Aznauryan}
     \affiliation{\JLAB}
     \affiliation{\YEREVAN}
\author{V.D.~Burkert}
     \affiliation{\JLAB}
\begin{abstract}
{
We utilize a light-front relativistic quark model 
to predict the $3q$ core contribution to the
electroexcitation of nucleon resonances of the $[70,1^-]$ multiplet
on the proton and neutron at $Q^2<5~$GeV$^2$. 
The investigation is stimulated in large degree by expected progress
in the studies of the electroexcitation of nucleon resonances
in the third resonance region in the CLAS experiment.
For the resonances $N(1520)\frac{3}{2}^-$, $N(1535)\frac{1}{2}^-$, and
$N(1675)\frac{5}{2}^-$, experimental data on electroexcitation amplitudes
on the proton are available in a wide range of $Q^2$. This allowed us to 
quantify the expected meson-baryon contributions to these
amplitudes as a function of $Q^2$. 
}
\end{abstract}
\pacs{ 12.39.Ki, 13.40.Gp, 14.20.Gk}
\maketitle

\section{Introduction}
\label{intro}
Experiments on the new generation of electron beam facilities
CEBAF(Jefferson Lab), MAMI(Mainz), and MIT-Bates
led to dramatic progress in the investigation of the
electroexcitation of nucleon resonances, and significant role
in the interpretation of new data belongs to quark models,
in particular, to light-front relativistic quark models (LF RQM).
The CLAS measurements at Jefferson Lab
made possible, for the first time,
the determination of the electroexcitation
amplitudes of the Roper resonance
$N(1440)\frac{1}{2}^+$ on the proton
in a wide range of photon virtuality up to $Q^2=4.5~$GeV$^2$ \cite{CLAS2009}. 
The comparison of these results
with the LF RQM predictions \cite{Capstick1995,Aznauryan2007}
was crucial for identification of the 
$N(1440)\frac{1}{2}^+$ 
as a predominantly 
radial excitation of a three-quark (3$q$)
ground state, with additional non-3-quark
contributions needed to describe the low
$Q^2$ behavior of the amplitudes.
The $\gamma^*p\rightarrow \Delta(1232)\frac{3}{2}^+$
transition amplitudes have been measured in a more wide range
of $Q^2$ ($0.06 \div 8~$GeV$^2$) 
\cite{CLAS2009,Stave,Sparveris,Mertz,Kunz,
Frolov,Vilano,KELLY}. 
The obtained data strongly
confirm the meson-cloud contribution as a source of the long-standing
descreapancy between the data and quark model predictions for the 
magnetic-dipole form factor of this transition, and the 'bare'
contribution to this form factor,
obtained within dynamical reaction model \cite{Sato2001,Lee2004,EBAC} 
is very close to the LF RQM predictions \cite{Riska2004,Aznauryan2015,Aznauryan2016}.
Above $2~$GeV$^2$, the  LF RQM  \cite{Aznauryan2015} reproduces
observed in experiment smallness of the ratio $R_{EM}$,
as well the negative sign and sharply
growing absolute value of the ratio $R_{SM}$ for the
$\gamma^*p\rightarrow \Delta(1232)\frac{3}{2}^+$ transition.
A very interesting conclusion was made from the results on the 
$\gamma^*p\rightarrow N(1675)\frac{5}{2}^-$ amplitudes extracted from
CLAS data \cite{CLAS2015}. A special feature of the resonance
$N(1675)\frac{5}{2}^-$ is the strong suppression of the transverse
helicity amplitudes for its excitation through quark transition
from the proton. This feature allowed one to draw conclusion
regarding the dominant strength of the meson-baryon contribution
to the
$\gamma^*p\rightarrow N(1675)\frac{5}{2}^-$ 
transverse helicity amplitudes 
\cite{Aznauryan2015_1} which is supported by the results
of the dynamical coupled-channels approach \cite{EBAC}.

Experiments on meson electroproduction on new electron beam facilities
have been performed 
on the proton target and, in the whole,
allowed extraction of the electroexcitation ampltudes for the resonances
$\Delta(1232)\frac{2}{2}^+$ 
\cite{CLAS2009,Stave,Sparveris,Mertz,Kunz,
Frolov,Vilano,KELLY}
and 
$N(1535)\frac{1}{2}^-$ 
\cite{CLAS2009,Thompson,Denizli,Armstrong,Dalton} 
in the range of $Q^2$ up to $8~$GeV$^2$, for the
$N(1440)\frac{1}{2}^+$, 
$N(1520)\frac{3}{2}^-$, 
$N(1675)\frac{5}{2}^-$, 
$N(1680)\frac{5}{2}^+$, and 
$N(1710)\frac{1}{2}^+$ 
at $Q^2<4.5~$GeV$^2$
\cite{CLAS2009,CLAS2015,Mokeev1,Mokeev2,Mokeev3}, and for the
$\Delta(1620)\frac{1}{2}^-$, 
$N(1650)\frac{1}{2}^-$, 
$\Delta(1700)\frac{3}{2}^-$, and 
$N(1720)\frac{3}{2}^+$ at 
$Q^2=0.65 \div 1.3~$GeV$^2$
\cite{Mokeev1,Mokeev2,Mokeev3}. 
Currently new data are in preparation by the CLAS collaboration
for the $ep \to ep\pi^0$ process in the same kinematics region as the CLAS data
in the $ep \to e n \pi^+$ channel \cite{CLAS2015,CLAS2008}.
The two-channel analysis will allow for the separation of all resonances
in the third nucleon resonance region at
$Q^2<4.5~$GeV$^2$. Other processes, such as $en(p_s) \to ep\pi^-(p_s)$ 
on deuterium target and
$ep \to ep\pi^+\pi^-$ are also in preparation. 

Therefore, in the near future CLAS experiment will provide
us with rich information on the electroexcitation of the
nucleon resonances from the multiplet $[70,1^-]$
at $Q^2<4.5~$GeV$^2$, and our goal in the present investigation is to 
extend our previous results on the electroexcitation
of the $N(1520)\frac{3}{2}^-$ and $N(1535)\frac{1}{2}^-$ 
within LF RQM \cite{Aznauryan2012} by
comprehensive investigation of electroexcitation of all resonances 
assigned to the $[70,1^-]$-plet on the proton and neutron.

We use an approach based on the LF
dynamics which presents the most suitable framework
for describing the transitions between relativistic
bound systems \cite{Drell,Terentiev,Brodsky}.
In early works by Berestetsky and Terent'ev \cite{Terentiev},
the approach was based on the construction
of the generators of the Poincar\'e group in the LF.
It was later formulated
in the infinite momentum frame (IMF) \cite{Terentiev1,Aznauryan1982}.
This allowed one to demonstrate more clearly
that diagrams which violate
impulse approximation, i.e. the diagrams
containing  vertices like
$\gamma^*\rightarrow q{\bar q}$, do not contribute.
The interpretation of results for the
$\gamma^* N\rightarrow N(N^*)$ transitions in terms
of the vertices
$N(N^*)\leftrightarrow 3q$
and corresponding wave functions became more evident.
A similar approach was developed and used in  the investigation
of electroexcitation of nucleon resonances in Ref. \cite{Capstick1995}
within LF Hamiltonian dynamics \cite{Keister}.
Both approaches use complete set of orthogonal wave functions 
that correspond to the classification
of the nucleon and nucleon resonances
within the group $SU(6)\times O(3)$ in the c.m.s. of constituent quarks.
It was shown in Ref. \cite{Aznauryan1982} 
that the wave functions of the system of quarks
in the IMF and in their c.m.s. are related through Melosh rotations 
of quark spin matrices \cite{Melosh}.
The same result was obtained in Ref. \cite{Capstick1995}
within LF Hamiltonian dynamics.

The paper is organized as follows.
In Sec. \ref{formalism} we present the LF RQM formalism to compute
the $\gamma^* N\rightarrow N^*$ transition amplitudes.
We specify the IMF where the LF RQM is built and the relations
between the 
$<N^*|J_{em}^{\mu}|N>$  matrix elements 
and the $N(N^*)\leftrightarrow 3q$ wave functions in this frame.
Further, the relations between these matrix elements
and the $\gamma^* N\rightarrow N^*$ form factors 
and transition helicity amplitudes are presented.
In Sec. \ref{mixing} we discuss the mixings of the states $N(1535)\frac{1}{2}^-$ and 
$N(1650)\frac{1}{2}^-$, and
$N(1520)\frac{3}{2}^-$ and $N(1700)\frac{3}{2}^-$. We discuss and present
the available information on the corresponding mixing angles.
The results are presented in Sec. \ref{results} and further summarized and discussed
in Sec. \ref{summary}.

\section{The $\gamma^* N\rightarrow N^*$ transition amplitudes
in LF RQM}
\label{formalism}
The $\gamma^* N\rightarrow N^*$
transition amplitudes
have been evaluated within the approach
of Ref. \cite{Aznauryan1982} where the
LF RQM
is formulated in the infinite momentum frame
chosen in such a way, that the initial hadron moves
along the $z$-axis with the momentum ${\rm P}\rightarrow \infty$,
the virtual photon momentum is
${\rm k}^{\mu}=\left(
\frac {M^2-m^2-\mathbf{Q}^2_{\perp}}{4{\rm P}},
\mathbf{Q}_{\perp},
-\frac {M^2-m^2-\mathbf{Q}^2_{\perp}}{4{\rm P}}\right)$,
the final hadron momentum is
${\rm P'=P+k}$, and ${\rm Q^2\equiv -k^2}=\mathbf{Q}_{\perp}^2$;
$m$ and $M$
are masses of the nucleon and resonance, respectively.
In this frame, the matrix elements of the electromagnetic current
for the $\gamma^* N\rightarrow N^*$ transition
have the form:
\begin{eqnarray}
&& <N^*,S'_z|J_{em}^{\mu}|N,S_z>|_
{{\rm P}\rightarrow\infty} \nonumber \\
&&=3eQ_a\int \Psi'^+({\rm p}'_a,{\rm p}'_b,{\rm p}'_c) \Gamma_a^\mu
\Psi({\rm p}_a,{\rm p}_b,{\rm p}_c) d\Gamma,
\label{eq:sec1}
\end{eqnarray}
where $S_z$ and $S'_z$ are the projections of the hadron
spins on the $z$-direction.
In Eq. (\ref{eq:sec1}), it is supposed that
the photon interacts with quark $a$ (the quarks
in hadrons are denoted by $a,b,c$),
$Q_a$
is the charge of this quark in units of $e$ ($e^2/4\pi=1/137$);
$\Psi$ and $\Psi'$ are wave functions
in the vertices $N(N^*)\leftrightarrow 3q$;
${\rm p}_i$ and ${\rm p}'_i$ ($i=a,b,c$) are the quark momenta
in IMF;
$d\Gamma$ is the phase space volume;
$\Gamma_a^\mu$ corresponds to the vertex of the quark interaction
with the photon:
\begin{eqnarray}
&& x_a\Gamma_a^{x}=2{\rm p}_{ax}+{\rm Q}_x+i{\rm Q}_y\sigma_z^{(a)},\\
\label{eq:sec2}
&& x_a\Gamma_a^{y}=2{\rm p}_{ay}+{\rm Q}_y-i{\rm Q}_x\sigma_z^{(a)},\\
\label{eq:sec3}
&& \Gamma_a^{0}=\Gamma_a^z=2{\rm P},
\label{eq:sec4}
\end{eqnarray} 
where $x_i$ ($i=a,b,c$) is the fraction of the initial hadron momentum carried
by the quark:
\begin{equation}
\mathbf{p}_{i}=x_i\mathbf{P}+\mathbf{q}_{i\perp},~
~~\sum\limits_{i} {\mathbf{q}_{i\perp}}=0,
~~\sum\limits_{i} {x_i}=1.
\label{eq:sec5}
\end{equation} 

The invariant mass 
of the system of initial quarks has the form:
\begin{equation}
M_0^2=\left(\sum\limits_{i} {{\rm p}_i}\right)^2=
\sum\limits_{i} {\frac{\mathbf{q}_{i\perp}^2+m_q^2}{x_i}},
\label{eq:sec6}
\end{equation}
$m_q$ is the quark mass. 

Now we define
the c.m.s. of initial quarks with the quark 
three-momenta
$\mathbf{q}_{i}~(i=a,b,c)$, where quark transverse momenta are given
by Eqs. (\ref{eq:sec5}), and the z-components
are defined as: 
\begin{eqnarray} 
&&{\rm q}_{iz}+\omega_i=M_0x_i, ~~~\omega_i=\sqrt{m_q^2+\mathbf{q}_{i}^2},
\label{eq:secc}\\
&&{\rm q}_{iz}=\frac{1}{2}\left(x_iM_0-\frac{m_q^2+\mathbf{q}_{i\perp}^2}{x_iM_0}\right),
\label{eq:secc1}\\
&&M_0=\sum\limits_{i}
{\omega_i},~~~\sum\limits_{i} \mathbf{q}_{i}=0.
\label{eq:sec7}
\end{eqnarray} 
For the final state quarks,
the quantities defined by Eqs. (\ref{eq:sec5}-\ref{eq:sec7})
are expressed through $\mathbf{P}'$,
${\rm p}'_i$, $\mathbf{q}'_{i}$,
and $M'_0$.

According to results of Ref. \cite{Aznauryan1982},
the wave function $\Psi$ in Eq. (\ref{eq:sec1}) is related
to the wave function in the c.m.s. of quarks
defined according to Eqs. (\ref{eq:sec5}-\ref{eq:sec7})
through Melosh matrices \cite{Melosh}:
\begin{equation}
\Psi=U^+({\rm p}_a)U^+({\rm p}_b)U^+({\rm p}_c)\Psi_{fss}
\Phi(\mathbf{q}_{a},\mathbf{q}_{b},\mathbf{q}_{c}).
\label{eq:sec8}
\end{equation}
Here we have separated the flavor-spin-space ($\Psi_{fss}$)
and spatial ($\Phi$)
parts of the c.m.s. wave function.
The Melosh matrices are
\begin{equation}
U({\rm p}_i)=\frac{m_q+M_0x_i+i\epsilon _{lm}\sigma_l {\rm q}_{im}}
{\sqrt{(m_q+M_0x_i)^2+\mathbf{q}_{i\perp}^2}}.
\label{eq:sec9}
\end{equation} 
We construct the flavor-spin-space
parts of the wave functions
in the c.m.s. of quarks
by utilizing
the rules \cite{Capstick1995,Isgur1}
that correspond to the classification
of the nucleon and nucleon resonances
within the group $SU(6)\times O(3)$.

The phase space volume in Eq. (\ref{eq:sec1})
has the form:   
\begin{equation}
d\Gamma= \frac{1}{(2\pi)^6}\frac
{d\mathbf{q}_{b\perp}d\mathbf{q}_{c\perp}dx_b dx_c}
{4x_ax_bx_c}.
\label{eq:sec10}
\end{equation}  

\subsection{The relations between matrix elements (\ref{eq:sec1})
and the $\gamma^* N\rightarrow N^*$ transition helicity amplitudes}
\label{relations}

Electroexcitation of the states with
$J^P=\frac{1}{2}^-$ and 
$J^P=\frac{3}{2}^-,\frac{5}{2}^-$, that enter the multiplet
$[70,1^-]$, is described, respectively,
by two and three form factors,
which we define according to
Refs. \cite{Aznauryan_review,Devenish} in the following way:

\begin{eqnarray}
&<N^*(\frac{1}{2}^-)|J_{em}^{\mu}|N>\equiv e\bar{u}(P')
\gamma_5\tilde{J}^{\mu}u(P),
\label{eq:sec11}\\
&<N^*(\frac{3}{2}^-)|J_{em}^{\mu}|N>\equiv e\bar{u}_{\nu}(P')\Gamma^{\nu\mu}
u(P),
\label{eq:sec12}\\
&<N^*(\frac{5}{2}^-)|J_{em}^{\mu}|N>\equiv e\bar{u}_{\nu\nu_1}(P')k^{\nu_1}
\gamma_5\Gamma^{\nu\mu}u(P),
\label{eq:sec13}
\end{eqnarray}
where
\begin{eqnarray}
&{\tilde{J}}^{\mu} =
\left(k\hspace{-1.8mm}\slash k^{\mu}-k^2\gamma^{\mu}\right)G_1
+\left[k\hspace{-1.8mm}\slash {\cal P}^{\mu}-({\cal P}k)\gamma^{\mu}\right]G_2,
\label{eq:sec14}\\
&\Gamma^{\nu\mu}(Q^2)= G_1{\cal H}_1^{\nu\mu}+
G_2{\cal H}_2^{\nu\mu}+G_3{\cal H}_3^{\nu\mu},
\label{eq:sec15}\\
&{\cal H}_1^{\nu\mu}=k\hspace{-1.8mm}\slash
g^{\nu\mu}-k^{\nu}\gamma^{\mu},
\label{eq:sec16}\\
&{\cal H}_2^{\nu\mu}=k^{\nu}P'^{\mu}-(kP')g^{\nu\mu},
\label{eq:sec17}\\
&{\cal H}_3^{\nu\mu}=k^{\nu}k^{\mu}-k^2g^{\nu\mu},
\label{eq:sec18}
\end{eqnarray}
${\cal P}\equiv \frac{1}{2}(P'+P)$, $u(P),u(P')$
are the Dirac spinors, and 
$u_{\nu}(P')$, $u_{\nu\nu_1}(P')$ 
are the generalized Rarita-Schwinger spinors.

In the LF RQM under consideration, the form factors $G_i(Q^2)$ are derived through 
the matrix elements (\ref{eq:sec1}). 
For the 
$J^P={\frac{1}{2}}^-$ resonances, the relations between
form factors and the matrix elements (\ref{eq:sec1})
are following: 

\begin{eqnarray}
&& \frac{1}{2P}<N^*,\frac{1}{2}|J_{em}^{0,z}|N,\frac{1}{2}>|_
{P\rightarrow\infty}=Q^2G_1(Q^2),
\label{eq:app1}
\\
&& \frac{1}{2P}<N^*,\frac{1}{2}|J_{em}^{0,z}|N,-\frac{1}{2}>|_
{P\rightarrow\infty}=\nonumber \\
&&~~~~~~~~~~~~~~=-\frac{M+m}{2}QG_2(Q^2).
\label{eq:app2}
\end{eqnarray}
For the 
$J^P={\frac{3}{2}}^-$ resonances, these relations 
are following: 
\begin{eqnarray}
\frac{1}{2P}<N^*,\frac{3}{2}|J_{em}^{0,z}|N,\frac{1}{2}>|_
{P\rightarrow\infty}=~~~~~~~~~~~~~&\nonumber \\
 -\frac{Q}{\sqrt{2}}
\left[G_1(Q^2)-\frac{M+m}{2}G_2(Q^2)\right],~~~~~~~~~~&
\label{eq:app3}
\\
 \frac{1}{2P}<N^*,\frac{3}{2}|J_{em}^{0,z}|N,-\frac{1}{2}>
|_{P\rightarrow\infty}=
\frac{Q^2}{2\sqrt{2}}G_2(Q^2),&
\label{eq:app4}
\\
<N^*,\frac{3}{2}|J_{em}^{x}+iJ_{em}^{y}|N,-\frac{1}{2}>
|_{ P\rightarrow\infty}=
\frac{Q^3}{\sqrt{2}}G_3(Q^2).&
\label{eq:app5}
\end{eqnarray}

For the 
$J^P={\frac{5}{2}}^-$ resonances, we have: 
\begin{eqnarray}
\frac{1}{2P}<N^*,\frac{5}{2}|J_{em}^{0,z}|N,\frac{1}{2}>|_
{P\rightarrow\infty}=~~~~~~~~~~~~~&\nonumber \\
 -Q^2
\left[G_1(Q^2)+\frac{M-m}{2}G_2(Q^2)\right],~~~~~~~~~~&
\label{eq:app6}
\\
 \frac{1}{2P}<N^*,\frac{5}{2}|J_{em}^{0,z}|N,-\frac{1}{2}>
|_{P\rightarrow\infty}=
-\frac{Q^3}{2}G_2(Q^2),&
\label{eq:app7}
\\
<N^*,\frac{5}{2}|J_{em}^{x}+iJ_{em}^{y}|N,-\frac{1}{2}>
|_{ P\rightarrow\infty}=
Q^4G_3(Q^2).&
\label{eq:app8}
\end{eqnarray}

The relations between
the $\gamma^* N\rightarrow N^*{\frac{1}{2}}^-$
helicity amplitudes and the form factors $G_1(Q^2),G_2(Q^2)$
are following:
\begin{eqnarray}
&&A_{\frac{1}{2}}=
b\left[2Q^2G_1-(M^2-m^2)G_2\right],
\label{eq:ap1}
\\
&&S_{\frac{1}{2}}=-
b\frac{K}{\sqrt{2}}{\tilde S}_{\frac{1}{2}},
\label{eq:ap2}
\\
&&{\tilde S}_{\frac{1}{2}}=2(M- m)G_1+
(M+ m)G_2,
\label{eq:ap3}
\\
&&b\equiv e\sqrt{\frac{Q_{+}}{8m(M^2-m^2)}},
\label{eq:ap4}
\\
&& K\equiv \frac{\sqrt{Q_{+}Q_{-}}}{2M},
\label{eq:ap5}
\\
&& Q_{\pm}\equiv (M \pm m)^2+Q^2.
\label{eq:ap6}
\end{eqnarray}

For the resonances with $J^P={\frac{3}{2}}^-$ and ${\frac{5}{2}}^-$ we have:

\begin{eqnarray}
&&{A}_{1/2}=h_3X,
~~~{ A}_{3/2}=\mp \sqrt{3}h_2X,
\label{eq:ap7}
\\
&&{S}_{1/2}=\mp h_1\frac{K}{\sqrt{2}M}X,
\label{eq:ap8}
\\
&&X\equiv K^{l-1}e\sqrt{\frac{Q_{\pm}}
{32Jm(M^2-m^2)}},~l=J-\frac{1}{2},
\label{eq:ap9}
\end{eqnarray}
where
\begin{eqnarray}
&&h_1(Q^2)=\mp 4MG_1(Q^2)+4M^2G_2(Q^2)+\nonumber \\
&&~~~~~~~~~~~2(M^2-m^2-Q^2)G_3(Q^2),
\label{eq:ap10}\\
&&h_2(Q^2)=-2(\mp M+ m)G_1(Q^2)-\nonumber \\
&&(M^2-m^2-Q^2)G_2(Q^2)+
2Q^2G_3(Q^2),
\label{eq:ap11}\\
&&h_3(Q^2)=\pm\frac{2}{M}[Q^2+m(\mp
M+m)]G_1(Q^2)+\nonumber \\
&&(M^2-m^2-Q^2)G_2(Q^2)-2Q^2G_3(Q^2),
\label{eq:ap12}
\end{eqnarray}
and the upper and lower signs correspond, respectively, to 
$J^P={\frac{3}{2}}^-$ and ${\frac{5}{2}}^-$ resonances.

\section{Mixing of $N(1535)\frac{1}{2}^-$, $N(1650)\frac{1}{2}^-$, and
$N(1520)\frac{3}{2}^-$, $N(1700)\frac{3}{2}^-$ }
\label{mixing}

The multiplet $[70,1^-]$ consists of the following states: 
$N\frac{1}{2}^-(^28_{1/2})$, 
$N\frac{3}{2}^-(^28_{3/2})$, 
$N\frac{1}{2}^-(^48_{1/2})$, 
$N\frac{3}{2}^-(^48_{3/2})$, 
$N\frac{5}{2}^-(^48_{5/2})$, 
$\Delta\frac{1}{2}^-(^210_{1/2})$, and 
$\Delta\frac{3}{2}^-(^210_{3/2})$, where we use the notation
$^{2S+1}SU(3)_J$, which gives the assignment of the state
according to the $SU(3)$ group, $S$ is the total spin of the quarks,
and $J$ is the spin of the resonance.
The resonances with $J^P=\frac{1}{2}^-$ and $\frac{3}{2}^-$
can be composed, respectively, from the states $^28_{1/2},~^48_{1/2}$ 
and $^28_{3/2},~^48_{3/2}$, and therefore can be mixings of these states:  

\begin{eqnarray}
&N(1535)\frac{1}{2}^-=cos\theta_S|^28_{1/2}>-sin\theta_S|^48_{1/2}>,
\label{eq:mix1}\\
&N(1650)\frac{1}{2}^-=sin\theta_S|^28_{1/2}>+cos\theta_S|^48_{1/2}>,
\label{eq:mix2}\\
&N(1520)\frac{3}{2}^-=cos\theta_D|^28_{3/2}>-sin\theta_D|^48_{3/2}>,
\label{eq:mix3}\\
&N(1700)\frac{3}{2}^-=sin\theta_D|^28_{3/2}>+cos\theta_D|^48_{3/2}>.
\label{eq:mix4}
\end{eqnarray}
There is information on the mixing angles $\theta_S$ and $\theta_D$,
obtained from the description of resonance masses within quark model 
with QCD-inspired
interquark forces \cite{Isgur2} and from
experimental data on
the decay widths of the resonances in the $\pi N$ channel \cite{Hey}. 
The results of Ref. \cite{Hey}
are based on the relations:
\begin{eqnarray}
&<\pi N|^28_{1/2}>/<\pi N|^48_{1/2}>=-2,
\label{eq:mix5}\\
&<\pi N|^28_{3/2}>/<\pi N|^48_{3/2}>=2\sqrt{10},
\label{eq:mix6}
\end{eqnarray}
that follow from the $SU(6)_W$-symmetry. The same relations have been obtained
in Ref. \cite{Aznauryan1985} within the LF RQM by relating the
$<\pi N|N^*>$ amplitudes to the matrix elements of the axial-vector current 
$<N^*|J^{\mu}_{ax}|N>$ using the hypothesis of partially conserved
axial-vector current (PCAC) in the way suggested in Ref. \cite{PCAC}.
The results of Ref. \cite{Hey} are based on early data.
Using recent data \cite{RPP}, 
we have revised the values of the mixing
angles extracted from the $\pi N$ widths of the resonances.
As a result, we have obtained
\begin{equation}
\theta_S=-16.6\pm 5^{\circ},~~~~\theta_D=11.5\pm 4^{\circ},
\label{eq:mix7}
\end{equation}
instead of $\theta_S=-31.9^{\circ}$ and $\theta_D=10.4^{\circ}$ in Ref. \cite{Hey}.
Large difference in $\theta_S$ is caused mainly by the significant
change of the $N(1535)\frac{1}{2}^-\rightarrow \pi N$ width, that resulted
in increasing of
the ratio of the mean values of the 
$N(1535)\frac{1}{2}^-$ and $N(1650)\frac{1}{2}^-$ 
$\pi N$ decay widths from 0.3 to 0.8. 

The mixing angles obtained from the description of masses \cite{Isgur2} 
are following:
\begin{equation}
\theta_S=-32^{\circ},~~~~\theta_D=6.3^{\circ}.
\label{eq:mix8}
\end{equation}

\section{Results }
\label{results}

\begin{figure*}[htp]
\begin{center}
\includegraphics[width=12.0cm]{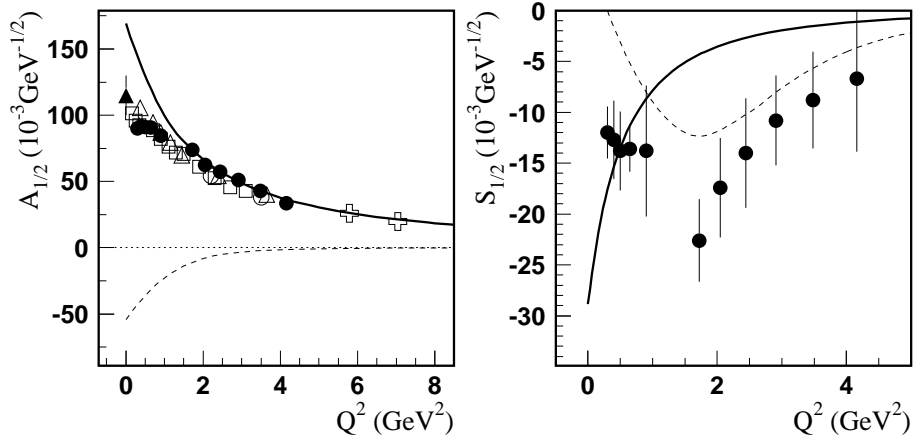}
\vspace{-0.1cm}
\caption{\small
The $\gamma^*p\rightarrow N(1535)\frac{1}{2}^-$
transition helicity amplitudes.
The solid curves are the LF RQM predictions;
the weight factors for the $3q$ contributions to the nucleon and
resonance are taken into account according to
Eqs. (\ref{eq:res1}) and (\ref{eq:res6}) with
$c_{N^*}=0.84$ and $0.94$ for the mixing angles
$\theta_S=-16.6^{\circ}$ and $-32^{\circ}$, respectively, 
(see Eqs.(\ref{eq:mix7}) and (\ref{eq:mix8})).
The thin dashed curves present the inferred meson-baryon 
contributions (see Sec. \ref{results3}).
Solid circles are the amplitudes extracted
from CLAS pion electroproduction data \cite{CLAS2009}.
The open triangles \cite{Thompson} and
open boxes \cite{Denizli}
are the amplitudes extracted from 
the JLab/Hall B $\eta$ electroproduction data;
the open circles \cite{Armstrong}
and open crosses \cite{Dalton}
are the amplitudes extracted from 
the JLab/Hall C $\eta$ electroproduction data;
the full triangle at $Q^2=0$ is the RPP 
estimate \cite{RPP}.
\label{fig:s11_1_p}}
\end{center}
\end{figure*}

\begin{figure*}[htp]
\begin{center}
\includegraphics[width=12.0cm]{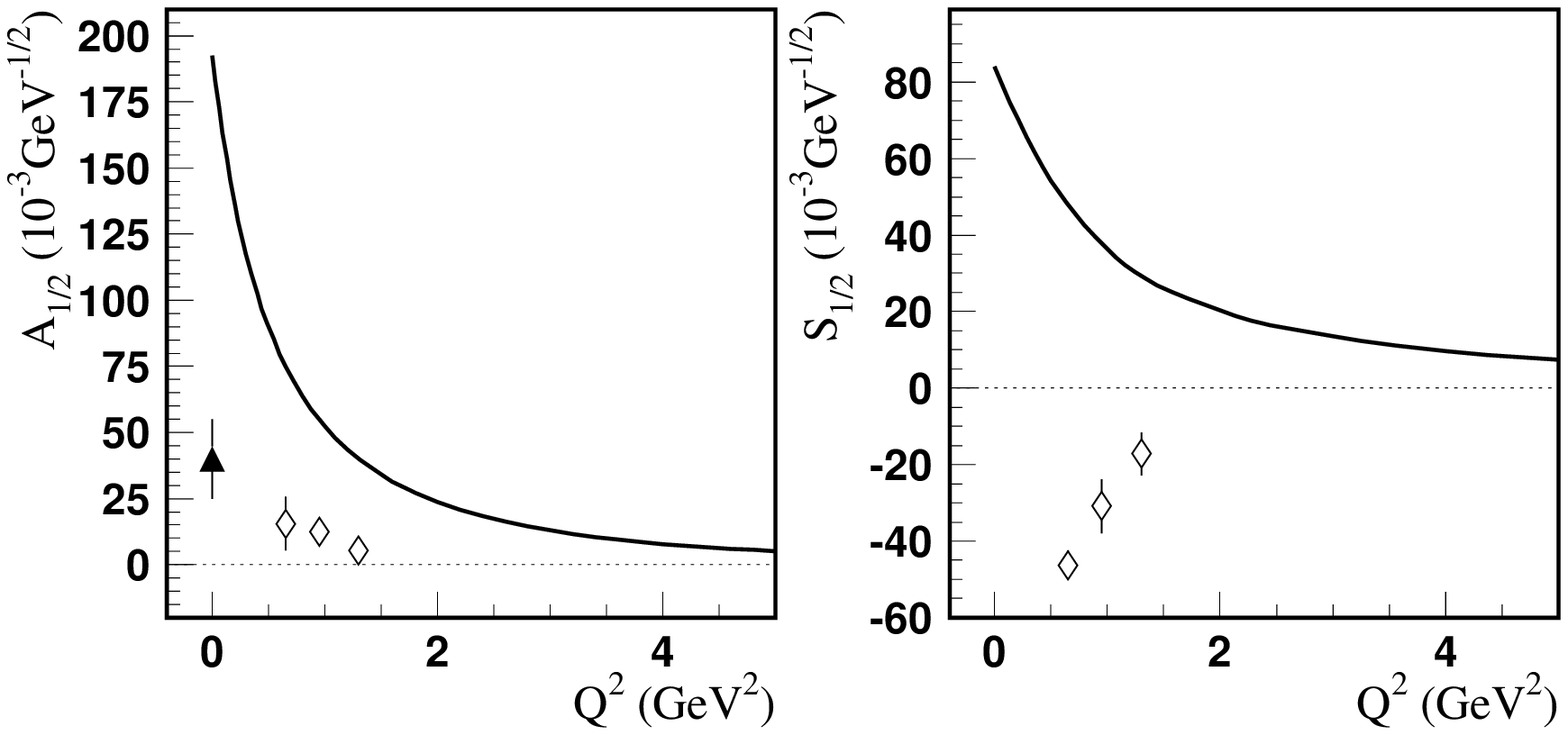}
\vspace{-0.1cm}
\caption{\small
The $\gamma^*p\rightarrow \Delta(1620)\frac{1}{2}^-$
transition helicity amplitudes.
The solid curves are the LF RQM predictions.
The full triangle at $Q^2=0$ is the RPP estimate
\cite{RPP};
open rhombuses
are the amplitudes extracted from
CLAS $2\pi$ electroproduction data \cite{Mokeev2}.
\label{fig:s31}}
\end{center}
\end{figure*}

\begin{figure*}[htp]
\begin{center}
\includegraphics[width=17.0cm]{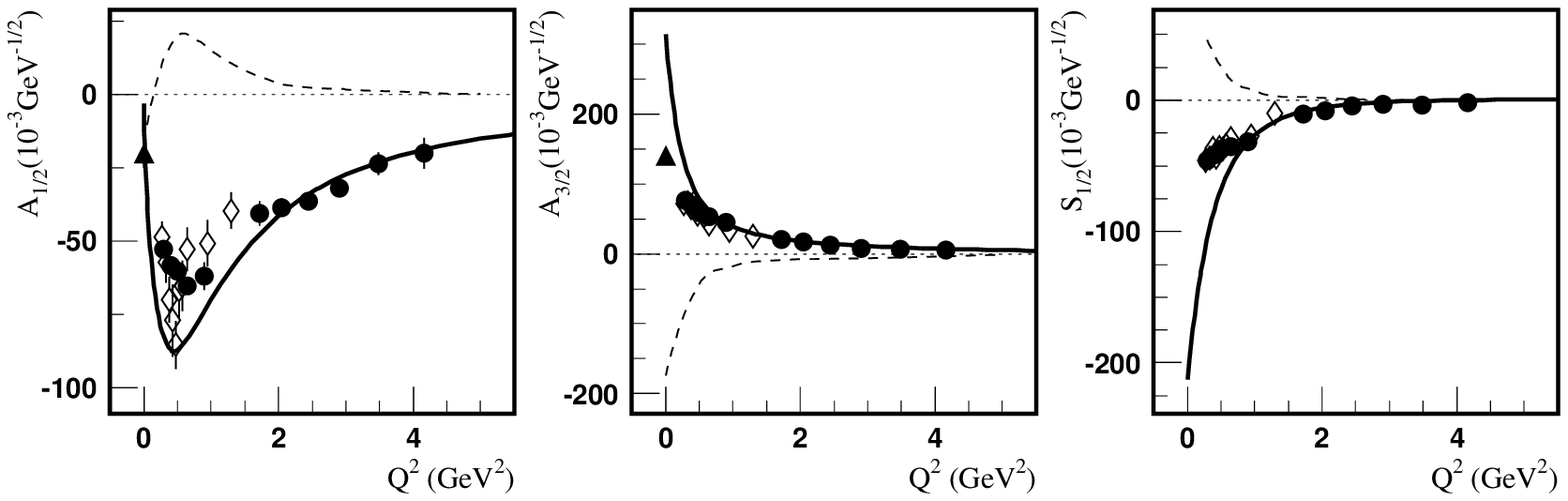}
\vspace{-0.1cm}
\caption{\small
The $\gamma^*p\rightarrow N(1520)\frac{3}{2}^-$
transition helicity amplitudes.
The solid curves are the LF RQM predictions;
the weight factors for the $3q$ contributions to the nucleon and
resonance are taken into account according to
Eqs. (\ref{eq:res1}) and (\ref{eq:res6}) with
$c_{N^*}=0.92$ and $0.94$ for the mixing angles
$\theta_D=6.3^{\circ}$ and $11.5^{\circ}$, respectively, 
(see Eqs.(\ref{eq:mix7}) and (\ref{eq:mix8})).
The thin dashed curves present the inferred meson-baryon 
contributions (see Sec. \ref{results3}).
Solid circles are the amplitudes extracted
from CLAS pion electroproduction data \cite{CLAS2009};
open rhombuses
are the amplitudes extracted from
CLAS $2\pi$ electroproduction data \cite{Mokeev1,Mokeev2,Mokeev3}.
The full triangles at $Q^2=0$ are the RPP estimates
\cite{RPP}.
\label{fig:d13_1_p}}
\end{center}
\end{figure*}

\begin{figure*}[htp]
\begin{center}
\includegraphics[width=17.0cm]{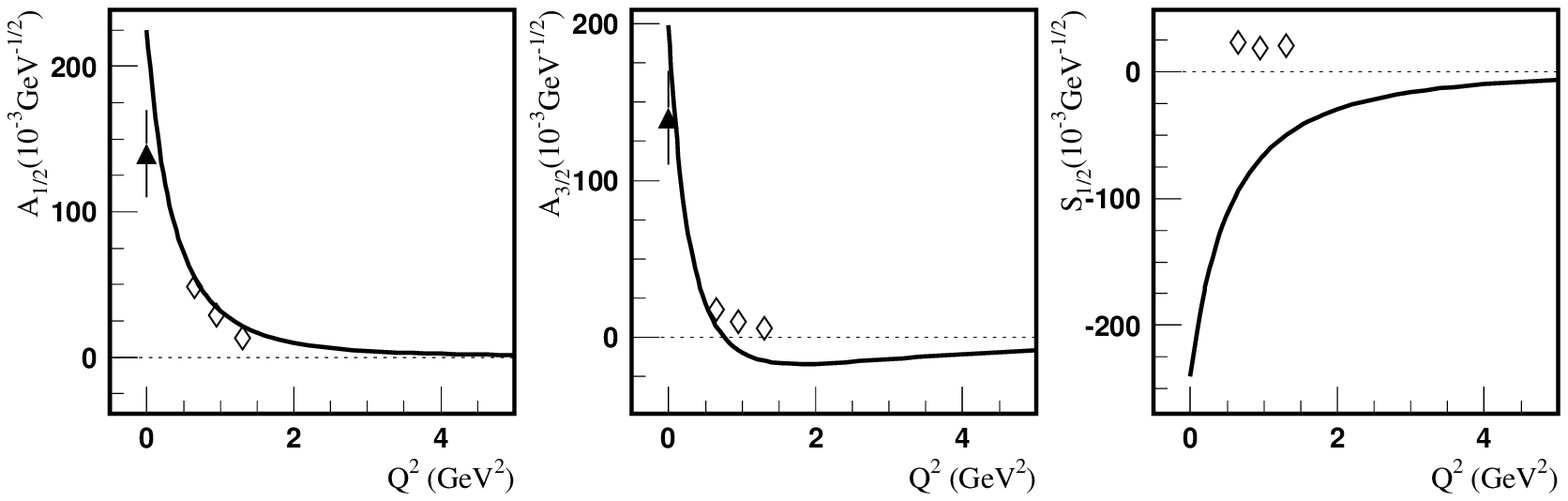}
\vspace{-0.1cm}
\caption{\small
The $\gamma^*p\rightarrow \Delta(1700)\frac{3}{2}^-$
transition helicity amplitudes.
The solid curves are the LF RQM predictions.
The full triangles at $Q^2=0$ are the RPP estimates
\cite{RPP};
open rhombuses
are the amplitudes extracted from
CLAS $2\pi$ electroproduction data \cite{Mokeev1}.
\label{fig:d33}}
\end{center}
\end{figure*}

\begin{figure*}[htp]
\begin{center}
\includegraphics[width=17.0cm]{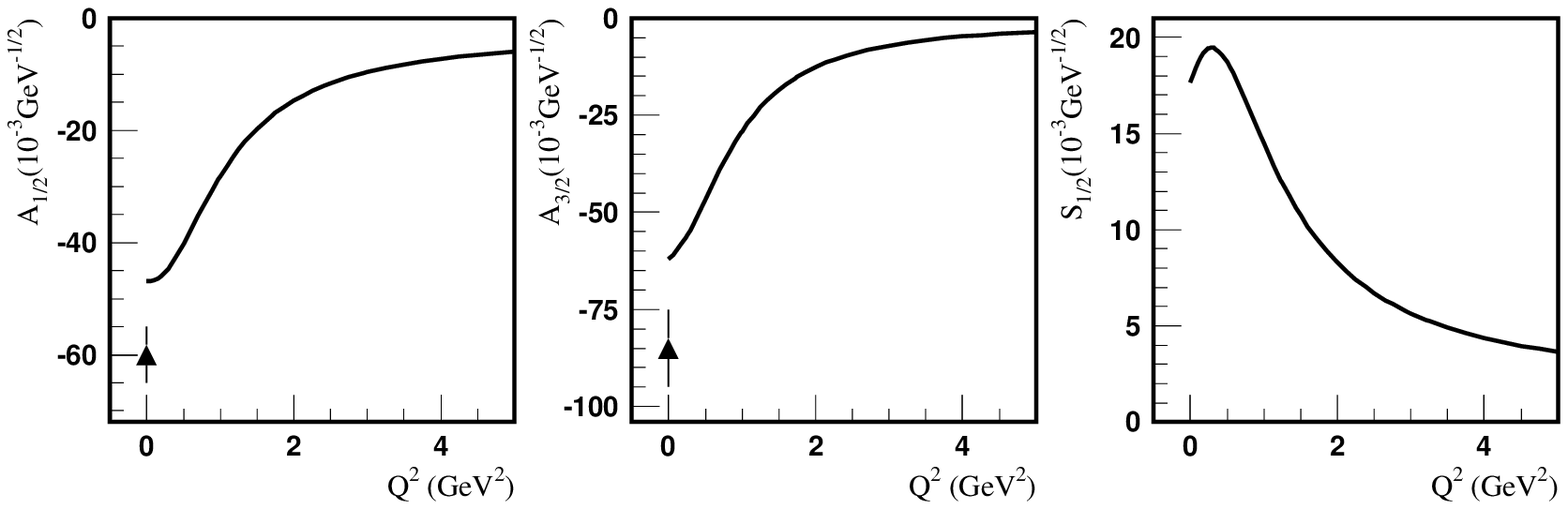}
\vspace{-0.1cm}
\caption{\small
The $\gamma^*n\rightarrow N(1675)\frac{5}{2}^-$
transition helicity amplitudes.
The solid curves are the LF RQM predictions.
The full triangles at $Q^2=0$ are the RPP estimates
\cite{RPP}.
\label{fig:d15_n}}
\end{center}
\end{figure*}

In this Section we present our results for the $3q$ core contribution
to the helicity transition amplitudes for the electroexcitation
of the resonances of the multiplet $[70,1^-]$ on the proton and neutron
(Figs. \ref{fig:s11_1_p}-\ref{fig:d15_n}).
The spacial part of the wave functions and parameters of the model
have been specified in Ref. \cite{Aznauryan2012} via description
of the nucleon electromagnetic form factors 
by combining $3q$ and pion-cloud contributions in the LF dynamics. 
Good description
of the nucleon electromagnetic form factors up to $Q^2=16~$GeV$^2$
has been obtained with the nucleon wave function in the form: 
\begin{equation}
|N>=0.95|3q>+0.313|\pi N>,
\label{eq:res1}
\end{equation}
and by employing two forms of the spatial wave function:
\begin{eqnarray}
&&\Phi_1 \sim exp(-M_0^2/6\alpha_1^2),
\label{eq:res2}
\\
&&\Phi_2 \sim
exp\left[-({\bf{q}}_a^2+{\bf{q}}_b^2+{\bf{q}}_c^2)/2\alpha_2^2\right],
\label{eq:res3}
\end{eqnarray}
with the following 
oscillator parameters and
running quark masses: 
\begin{eqnarray}
\alpha_1=0.37~{\rm GeV},~~
m_q^{(1)}(Q^2)=\frac{0.22{\rm GeV}}{1+Q^2/56{\rm GeV}^2},
\label{eq:res4}\\
\alpha_2=0.41~{\rm GeV},~~
m_q^{(2)}(Q^2)=\frac{0.22{\rm GeV}}{1+Q^2/18{\rm GeV}^2}.
\label{eq:res5}
\end{eqnarray} 
For the resonances of the $[70,1^-]$-plet, the results for
the transition amplitudes
obtained with the wave functions
(\ref{eq:res2},\ref{eq:res3}) 
and corresponding parameters
(\ref{eq:res4},\ref{eq:res5})
are very close
to each other. The role of running quark mass becomes visible
above $3~$GeV$^2$. 
At $Q^2=5~$GeV$^2$, 
it increases the transition helicity amplitudes
by $25-35\%$ and $10-15\%$ 
for the  wave functions (\ref{eq:res2}) and 
(\ref{eq:res3}), respectively.

Meson electroproduction gives strong evidence,
that baryon resonances are not excited from quark transition alone,
but there can be significant contribution 
from meson-baryon interaction, including pion-loop
contributions generated by nearly massles pions.
The common feature of all approaches
that account for meson-baryon contributions is the fact that
they are more rapidly losing their strength when $Q^2$
increases in comparison to the $3q$ contributions. For the
 $N(1535)\frac{1}{2}^-$ and $N(1520)\frac{3}{2}^-$,
it is expected, that meson-baryon contributions can be neglected
at $Q^2>2~$GeV$^2$ \cite{EBAC}. 
There are accurate data for the electroexcitation of 
these resonances
on the proton, respectively, at $Q^2<8$ and $4.5~$GeV$^2$.
Therefore, the weight of the $3q$ contributions to
the $N(1535)\frac{1}{2}^-$ and $N(1520)\frac{3}{2}^-$:
\begin{equation}
|N^*>=c_{N^*}|3q>+...,~~~~c_{N^*}<1,
\label{eq:res6}
\end{equation}
we find from experimental values of the transition helicity amplitudes, 
assuming that at $Q^2>2~$GeV$^2$ they are dominated by the $3q$ contributions.
The weight factors $c_{N^*}$ for the 
$N(1535)\frac{1}{2}^-$ and $N(1520)\frac{3}{2}^-$
are presented in the Captions to Figs. \ref{fig:s11_1_p}
and \ref{fig:d13_1_p}.

\subsection{Mixings and the results for the
$N(1535)\frac{1}{2}^-$, $N(1650)\frac{1}{2}^-$ and 
$N(1520)\frac{3}{2}^-$, $N(1700)\frac{3}{2}^-$}
\label{results1}

\begin{figure*}[htp]
\begin{center}
\includegraphics[width=12.0cm]{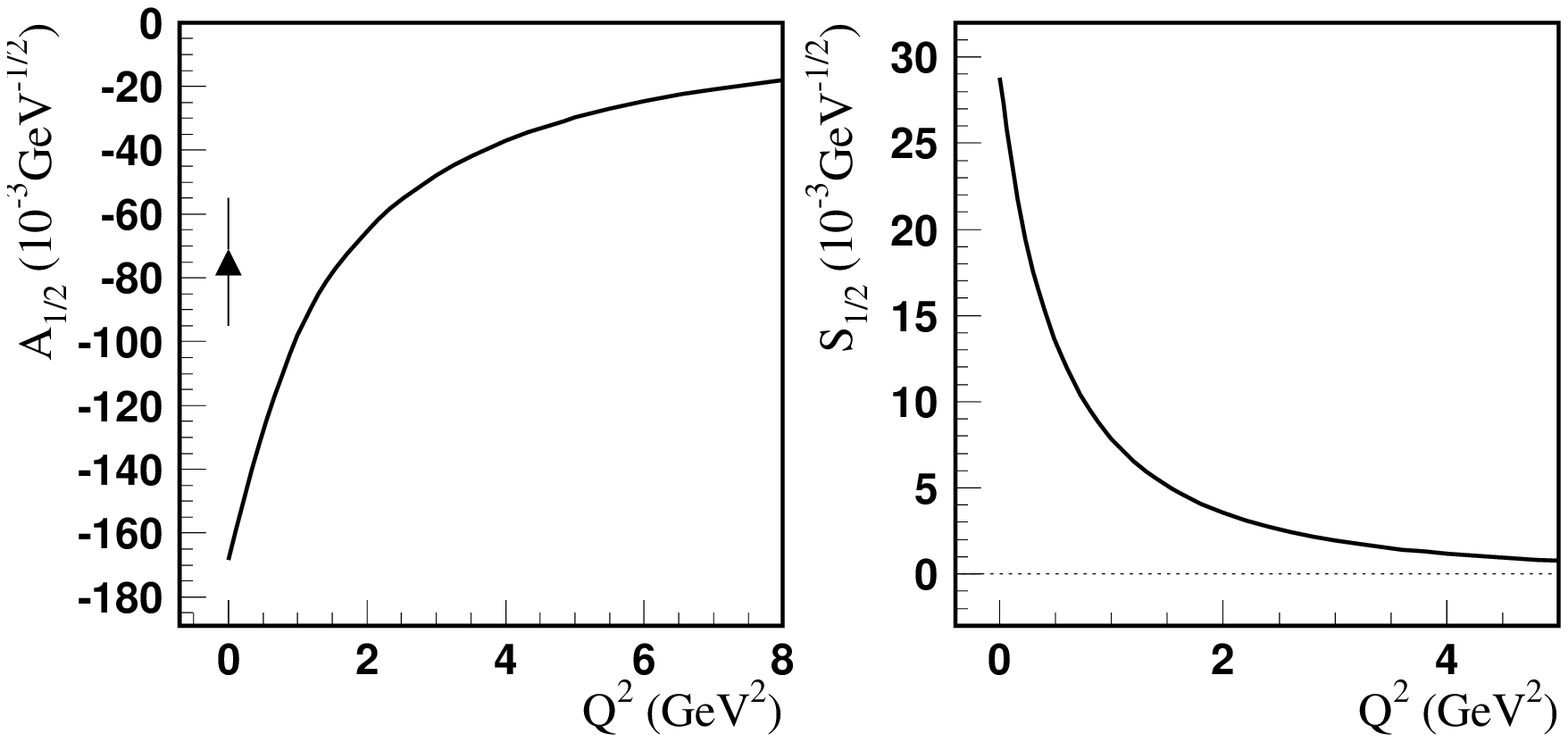}
\vspace{-0.1cm}
\caption{\small
The $\gamma^*n\rightarrow N(1535)\frac{1}{2}^-$
transition helicity amplitudes.
Legend for the solid curves is as for Fig. \ref{fig:s11_1_p}.
The full triangle at $Q^2=0$ is the RPP estimate
\cite{RPP}. 
\label{fig:s11_1_n}}
\end{center}
\end{figure*}

\begin{figure*}[htp]
\begin{center}
\includegraphics[width=12.0cm]{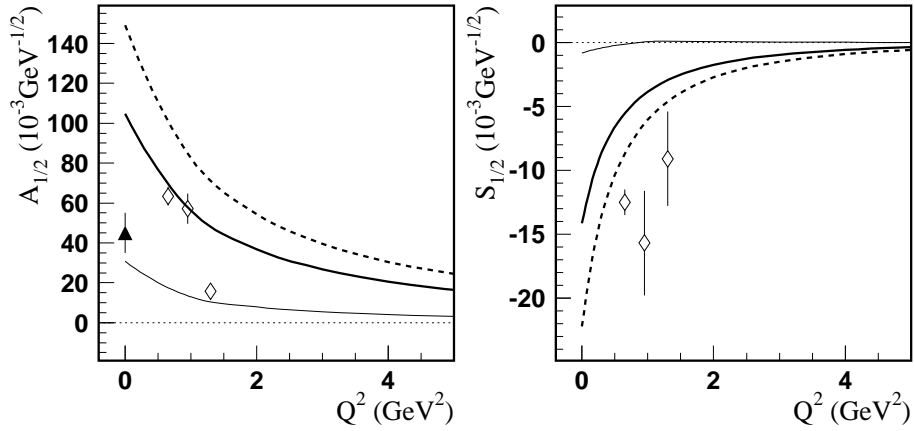}
\vspace{-0.1cm}
\caption{\small
The $\gamma^*p\rightarrow N(1650)\frac{1}{2}^-$
transition helicity amplitudes.
The LF RQM predictions
are shown by
thin and thick solid lines for the mixing angles
$\theta_S=0$ and $-16.6^{\circ}$, respectively,
and by thick dashed lines for 
$\theta_S=-32^{\circ}$
(see Eqs.(\ref{eq:mix7}) and (\ref{eq:mix8})).
The full triangle at $Q^2=0$ is the RPP 
estimate \cite{RPP};
open rhombuses
are the amplitudes extracted from
CLAS $2\pi$ electroproduction data \cite{Mokeev1}.
\label{fig:s11_2_p}}
\end{center}
\end{figure*}

\begin{figure*}[htp]
\begin{center}
\includegraphics[width=12.0cm]{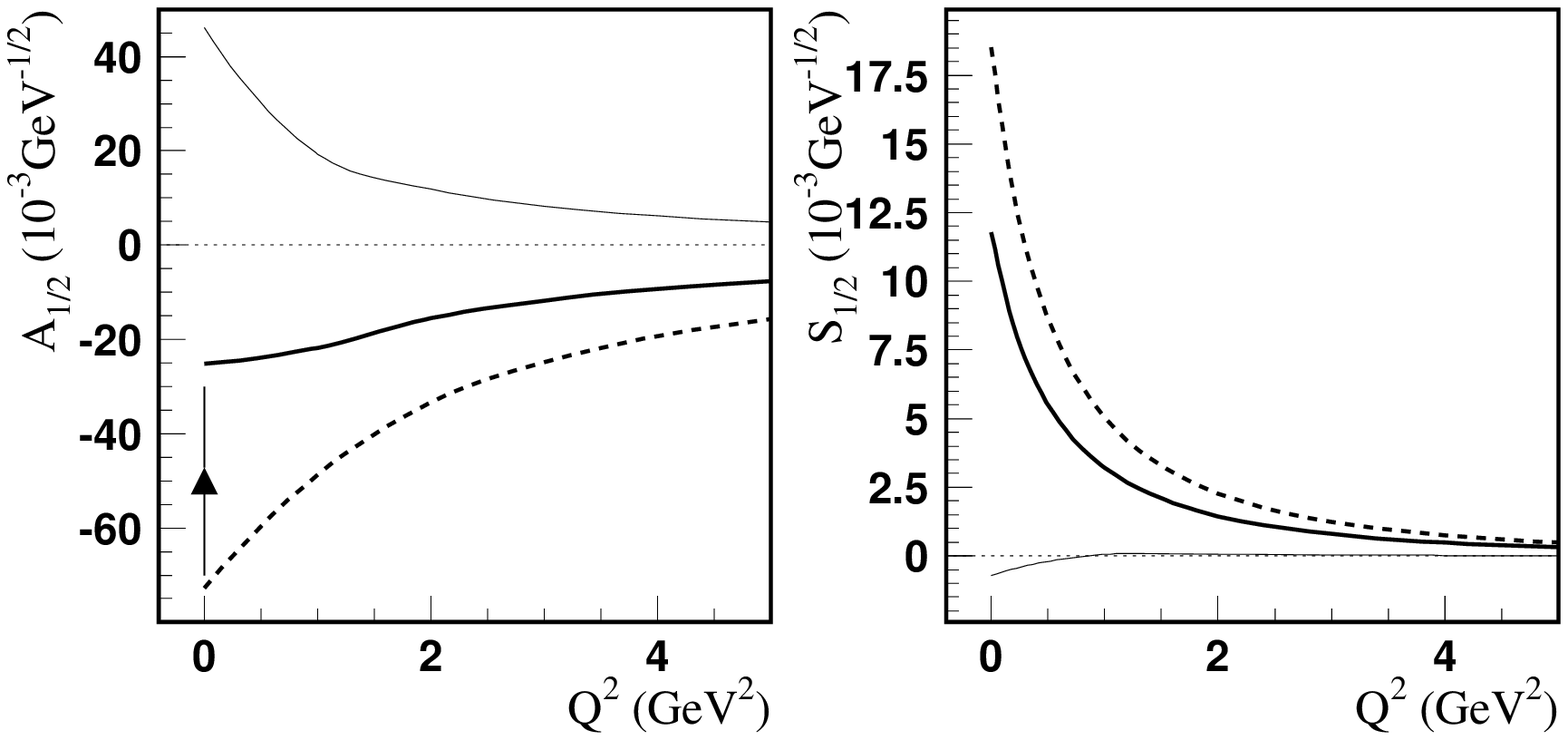}
\vspace{-0.1cm}
\caption{\small
The $\gamma^*n\rightarrow N(1650)\frac{1}{2}^-$
transition helicity amplitudes.
Legend for the lines is as for Fig. \ref{fig:s11_2_p}.
The full triangle at $Q^2=0$ is the RPP estimate
\cite{RPP}. 
\label{fig:s11_2_n}}
\end{center}
\end{figure*}

\begin{figure*}[htp]
\begin{center}
\includegraphics[width=17.0cm]{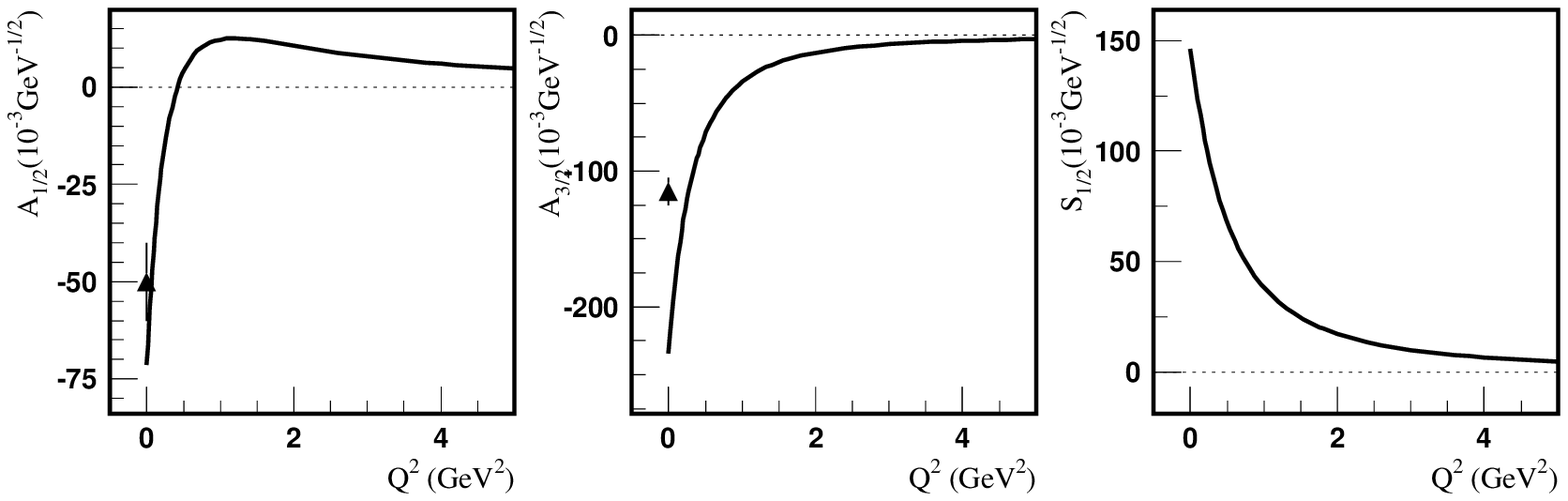}
\vspace{-0.1cm}
\caption{\small
The $\gamma^*n\rightarrow N(1520)\frac{3}{2}^-$
transition helicity amplitudes.
Legend for the solid curves is as for Fig. \ref{fig:d13_1_p}.
The full triangles at $Q^2=0$ are the RPP estimate
\cite{RPP}.
\label{fig:d13_1_n}}
\end{center}
\end{figure*}

\begin{figure*}[htp]
\begin{center}
\includegraphics[width=17.0cm]{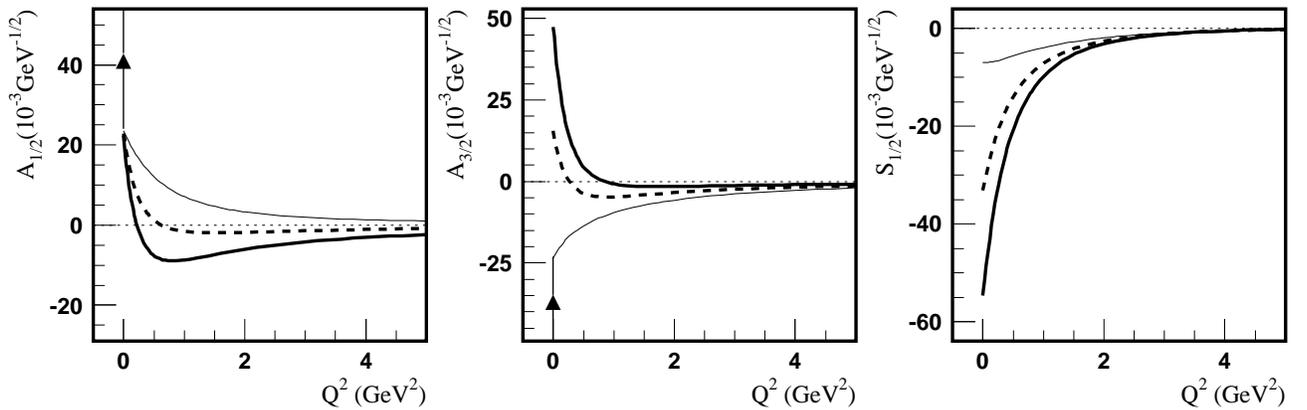}
\vspace{-0.1cm}
\caption{\small
The $\gamma^*p\rightarrow N(1700)\frac{3}{2}^-$
transition helicity amplitudes.
The LF RQM predictions are shown by the
thin and thick solid lines for the mixing angles
$\theta_S=0$ and $11.5^{\circ}$, respectively,
and by thick dashed lines for 
$\theta_S=6.3^{\circ}$
(see Eqs.(\ref{eq:mix7}) and (\ref{eq:mix8})).
The full triangles at $Q^2=0$ are the RPP estimates
\cite{RPP}. 
\label{fig:d13_2_p}}
\end{center}
\end{figure*}

\begin{figure*}[htp]
\begin{center}
\includegraphics[width=17.0cm]{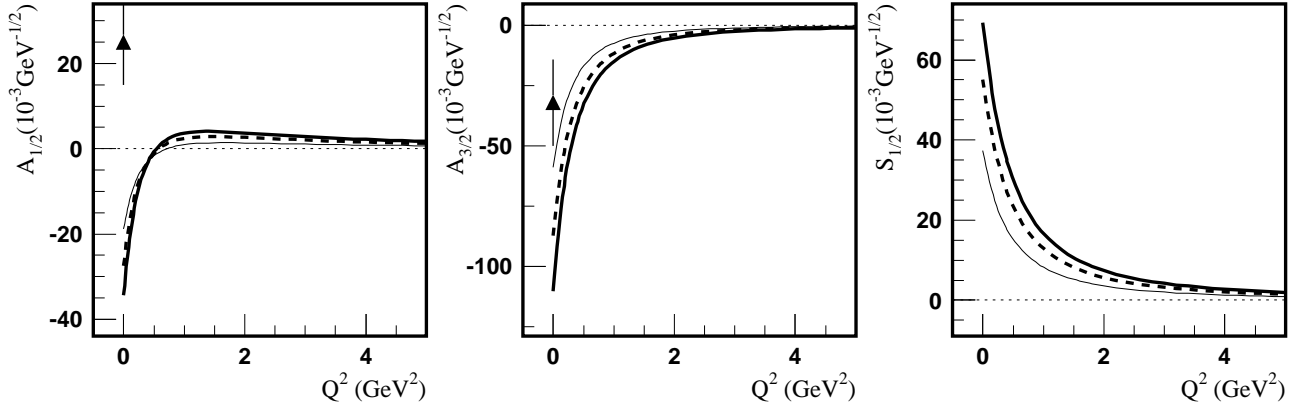}
\vspace{-0.1cm}
\caption{\small
The $\gamma^*n\rightarrow N(1700)\frac{3}{2}^-$
transition helicity amplitudes.
Legend for the lines and data is as for Fig. \ref{fig:d13_2_p}.
\label{fig:d13_2_n}}
\end{center}
\end{figure*}

The results for the resonances $N(1535)\frac{1}{2}^-$, $N(1650)\frac{1}{2}^-$
and $N(1520)\frac{3}{2}^-$, $N(1700)\frac{3}{2}^-$ are shown
in Figs. \ref{fig:s11_1_p}-\ref{fig:s11_2_n}
and \ref{fig:d13_1_p}-\ref{fig:d13_2_n}
taking into account mixings discussed in Section \ref{mixing}.
It can be seen, that the amplitudes
for the resonances $N(1650)\frac{1}{2}^-$ 
and $N(1700)\frac{3}{2}^-$, taken as pure $^4 8_{1/2}$
and $^4 8_{3/2}$ states, are significantly smaller than the
amplitudes for the 
$N(1535)\frac{1}{2}^-$ and $N(1520)\frac{3}{2}^-$. For this
reason, the mixings play significant role in the electroexcitation
of the $N(1650)\frac{1}{2}^-$ 
and $N(1700)\frac{3}{2}^-$, and    
in Figs. \ref{fig:s11_2_p},\ref{fig:s11_2_n} and
\ref{fig:d13_2_p},\ref{fig:d13_2_n},
we present three kind of curves: thin solid curves for the unmixed states
($\theta_S=\theta_D=0$) and thick solid and dashed curves,
respectively, for mixing angles from Eqs. (\ref{eq:mix7}) and (\ref{eq:mix8}).
For the resonances $N(1535)\frac{1}{2}^-$ 
and $N(1520)\frac{3}{2}^-$, the corresponding curves 
are very close to each other. 

It is known, that the results for the
$\gamma^* N\rightarrow N^*$ transition amplitudes extracted from
experimental data contain an additional sign related to the
vertex of the resonance coupling to the final state hadrons
(see, for example, Ref. \cite{Aznauryan_review}).  
In the electroproduction of pions on nucleons this is the relative sign
between the $\pi N N^*$ and $\pi N N$ vertices.
For the resonances of $[70,1^-]$-plet, this sign has been found
in Ref. \cite{Aznauryan1985} in the LF approach based on PCAC 
(see also Section \ref{mixing}).
In Ref. \cite{Aznauryan1985},
the electroexcitation of the resonances 
of $[70,1^-]$-plet on the proton and neutron has been investigated at $Q^2=0$, 
and the results for the transverse transition helicity amplitudes
have been presented taking into account the relative sign
between the $\pi N N^*$ and $\pi N N$ vertices.
This sign is taken into account also in
the results obtained in the present investigation
and shown in Figs. \ref{fig:s11_1_p}-\ref{fig:d15_n}.
We mention, that
from the relations (\ref{eq:mix1},\ref{eq:mix2},\ref{eq:mix5})
it follows that 
in all considered cases of mixings, the relative sign between
the $\pi N N(1535)$ and $\pi N N(1650)$ vertices is negative.
This is important for understanding of the results
for the $N(1650)\frac{1}{2}^{-}$.

\subsection{SQTM and the results for the
$N(1675)\frac{1}{2}^-$}
\label{results2}

\begin{figure*}[htp]
\begin{center}
\includegraphics[width=17.0cm]{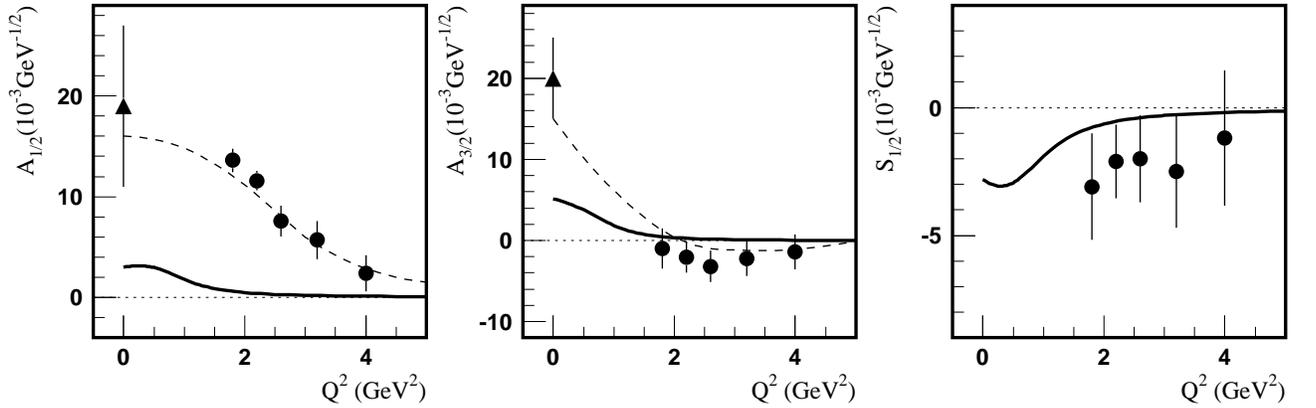}
\vspace{-0.1cm}
\caption{\small
The $\gamma^*p\rightarrow N(1675)\frac{5}{2}^-$
transition helicity amplitudes.
The solid curves are the LF RQM predictions.
The thin dashed curves present the inferred meson-baryon 
contributions (see Sec. \ref{results3}).
The full triangles at $Q^2=0$ are the RPP estimates
\cite{RPP};
the solid circles
are the amplitudes extracted from
CLAS $\pi$ electroproduction data \cite{CLAS2015}.
\label{fig:d15_p}}
\end{center}
\end{figure*}

Now we comment on the results for the $N(1675)\frac{5}{2}^-$,  
Figs. \ref{fig:d15_p},\ref{fig:d15_n}. The approximation 
of the single quark transition model (SQTM) 
\cite{Hey_Weyers,Babcock_Rosner,Cottingham,SQTM}
leads to selection rules, which for 
the resonances of the $[70,1^-]$-plet 
result in the suppression of the transition from the proton
to the states with $S=\frac{3}{2}$ for the transverse
helicity amplitudes. These are the states
$N\frac{1}{2}^-(^48_{1/2})$,
$N\frac{3}{2}^-(^48_{3/2})$,
and $N\frac{5}{2}^-(^48_{5/2})$. According to our results,
relativistic effects violate this suppression weakly.
For the $J=\frac{1}{2}$ and $\frac{3}{2}$ states,
this can be seen from 
Figs. \ref{fig:s11_2_p}, \ref{fig:d13_2_p}, where the amplitudes
for the electroexcitation of 
$N\frac{1}{2}^-(^48_{1/2})$
and $N\frac{3}{2}^-(^48_{3/2})$ are given by the thin
solid lines. 
For the resonance $N(1675)\frac{5}{2}^-$,
we also have small violation of the suppression of the
transverse helicity amplitudes for the electroexcitation on
the proton (see Fig. \ref{fig:d15_p}). 
In contrast with proton, electroexcitation amplitudes on the neutron
are large. In both cases, for proton and neutron,
close predictions have been
obtained in the quark model of Ref. \cite{Giannini}. 

\begin{table*}[t]
\begin{tabular}{|c|c|c|}
\hline
&&\\
Resonance&proton&neutron\\
&&\\
&$A_{1/2}~~~~~~~~~~A_{3/2}
~~~~~~~~~~A_{1/2}~~~~~~~~~~A_{3/2}$
&$A_{1/2}~~~~~~~~~~A_{3/2}
~~~~~~~~~A_{1/2}~~~~~~~~~~A_{3/2}$\\
&&\\
&$~~~~~$exp. \cite{RPP}$~~~~~~~~~~~~~~~~~$exp$~-~$LF RQM&$~~~~~$exp. \cite{RPP}$
~~~~~~~~~~~~~~~~~$exp$~-~$LF RQM\\
&&\\
\hline
&&\\
$N(1520){\frac{3}{2}}^-$&$-20\pm 5$~~~~~~~~$140\pm 10$ ~~~~$-17\pm 5$ ~~~ $-174\pm 10$
&$-50\pm 10$~~~$-115\pm 10$
~~~~$25\pm 10$~~~~$131\pm 10$\\
&&\\
$N(1535){\frac{1}{2}}^-$&$115\pm 15$~~~~~~~~~~~~~~~~~~~~~~~$-54\pm 15~~~~~~~~~~~~~~~~$
&$~-75\pm 20~~~~~~~~~~~~~~~~~~~~$ 
$102\pm 20~~~~~~~~~~~~~~~~~$\\
&&\\
$\Delta(1620){\frac{1}{2}}^-$&$40\pm 15$~~~~~~~~~~~~~~~~~~~~~~~$-152\pm 15~~~~~~~~~~~~~~~~$
&\\ 
&&\\
$N(1675){\frac{5}{2}}^-$&$19\pm 8$~~~~~~~~~~~~$20\pm 5$~~~~~~~~~$16\pm 8$ ~~~~~~~ $15\pm 5$
&$-60\pm 5$~~~$-85\pm 10$
~~~~$-13\pm 5$~~~~$-23\pm 10$\\
&&\\
$\Delta(1700){\frac{3}{2}}^-$&$140\pm 30$~~~~~~~~$140\pm 30$ ~~~~$-85\pm 30$ ~~~ $-59\pm 30$
&\\
&&\\
\hline
\end{tabular}
\caption{\label{cloud}
 Transverse transition helicity
amplitudes at $Q^2=0$ for several states of the $[70,1^-]$ multiplet 
for proton and neutron (in units of $10^{-3}GeV^{-1/2}$). 
The first two columns show the RPP estimates \cite{RPP}. 
Columns 3 and 4 show the inferred meson-baruon contributions obtained by
subtraction the values obtained in the LF RQM from experimental data. 
The quoted uncertainties are from the experimental estimates.  
\label{tab:cloud}}
\end{table*}

\subsection{Inferred meson-baryon contributions}
\label{results3}

For the resonances $N(1520)\frac{3}{2}^-$, $N(1535)\frac{1}{2}^-$, and
$N(1675)\frac{5}{2}^-$, experimental data on electroexcitation amplitudes
on the proton are available in wide range of $Q^2$. This allowed us to
quantify the expected meson-baryon contributions to these
amplitudes at $Q^2<2-3~$GeV$^2$.
The meson-baryon contributions inferred from the difference
of the LF RQM predictions and the data are shown 
in Figs. \ref{fig:s11_1_p}, \ref{fig:d13_1_p}, \ref{fig:d15_p}
by thin dashed lines. 
They correspond approximately to the mean values of experimental
data. The spread of these contributions can be deduced from the spread
and errors of experimental data. 

The constituent quark and inferred meson-baryon contributions
can be associated, respectively, with the bare and
meson-cloud contributions of the dynamical coupled-channels approaches
that incorporate hadronic and electromagnetic channels.
Much progress has been made recently within the EBAC/Argonne-Osaka 
coupled-channels analyses 
\cite{EBAC,EBAC1,EBAC2} 
that include pion photo- and electroproduction data.
However, 
only preliminary results are available
from the analyses that are based on the complete set
of the CLAS pion electroproduction data 
in the whole $Q^2$ range up to $4.5~$GeV$^2$
and from two channels
$ep \to ep\pi^0$ and $ep \to e n \pi^+$
\cite{INT,INT1}.
The results of the
coupled-channels analyses are related to the
resonance pole positions; with this in Refs. 
\cite{EBAC,EBAC1} the absolute values of the meson cloud contributions 
continued to the real axis  and evaluated at $W=1.535,~1.52$, and $1.625$ GeV,
respectively, for the
resonances $N(1520)\frac{3}{2}^-$, $N(1535)\frac{1}{2}^-$, and
$N(1675)\frac{5}{2}^-$ are presented.

All inferred meson-baryon contributions 
have clear peak at $Q^2=0$, except the contributions for the
$N(1520)\frac{3}{2}^-$ 
$A_{1/2}(Q^2)$ amplitude and for the 
$N(1535)\frac{1}{2}^-$ 
$S_{1/2}(Q^2)$ amplitude.
Such pronounced peak is specific 
for the corresponding meson cloud contributions in the
coupled-channels analyses 
\cite{EBAC,EBAC1,INT,INT1}.
Concerning  the $A_{1/2}(Q^2)$ amplitude for 
the $N(1520)\frac{3}{2}^-$, we mention that in all
coupled-channels analyses 
\cite{EBAC,EBAC1,INT,INT1}
the results for the meson
cloud contribution are by order of magnitude
and $Q^2$ dependence very close to our result. 

For the states that are not affected
by mixings, we present also in Table \ref{cloud}
the inferred meson-baryon contributions to 
the transverse transition helicity amplitudes at the photon point $Q^2=0$.
According to our results, these contributions for the 
$N(1520)\frac{3}{2}^-$, $N(1535)\frac{1}{2}^-$, and
$N(1675)\frac{5}{2}^-$ are dominated by the isovector component.

\section{Summary and discussion}
\label{summary}

In this paper we present the results of a comprehensive investigation
of electroexcitation of nucleon resonances of the multiplet
$[70,1^-]$ on the proton and neutron within LF RQM.
The investigation was stimulated by the expected progress in 
the extraction of the electroexcitation amplitudes for these resonances
from the CLAS data, and also by the experiments on deuterium target.

It is known, that the three-quark structure of baryons resulted
in predictions of a wealth of excited states with underlying spin-flavor
and orbital symmetry of $SU(6)\times O(3)$. In spite of the 
essentially non-relativistic nature of this symmetry, it describes well
the observed
quantum numbers and in many cases masses of the resonances 
in the first, second, and third nucleon resonance regions. 
The LF dynamics is known as most suitable framework for describing
transitions of baryons composed of relativistic
constituent quarks. The important feature of the LF approach of 
Ref. \cite{Aznauryan1982}, employed in the present investigation,
as well of the LF approach of
Ref. \cite{Capstick1995}, 
is the fact that these approaches could solve in uniform way 
the problem
of construction of orthogonal set of wave functions for 
the relativistic quarks by preserving the $SU(6)\times O(3)$ symmetry.
This has been done by setting the $SU(6)\times O(3)$ symmetry
in the c.m.s. of constituent quarks defined by Eqs. (\ref{eq:sec5}-\ref{eq:sec7}).
Then it was shown, that in the IMF or LF framework,  
which are used for calculation of the transition
amplitudes, the flavour-spin-space part of wave functions are related
to the wave functions in c.m.s. of quarks by 
quark spin rotations given by the Melosh matrices. 
Therefore, in our calculations we employ the flavor-spin-space
parts of the wave functions that
in the c.m.s. of quarks
correspond to the classification
of states
within the group $SU(6)\times O(3)$.

The pairs of resonances  $N(1535)\frac{1}{2}^-$, $N(1650)\frac{1}{2}^-$ and
$N(1520)\frac{3}{2}^-$, $N(1700)\frac{3}{2}^-$ with the same spin-parity
can be composed, respectively, from the states $^28_{1/2},~^48_{1/2}$
and $^28_{3/2},~^48_{3/2}$. Therefore, they can be mixings of these states.
There is information on the mixing angles,
obtained from the description of resonance masses within quark model
with QCD-inspired
interquark forces \cite{Isgur2} and from
experimental data on
the decay widths of the resonances in the $\pi N$ channel \cite{Hey}.
The results of Ref. \cite{Hey} are based on the early data.
Using recent data \cite{RPP}, we have revised the values of the mixing
angles extracted from the $\pi N$ widths of the resonances.
In our calculations of the electroexcitation amplitudes for
the $N(1535)\frac{1}{2}^-$, $N(1650)\frac{1}{2}^-$,
$N(1520)\frac{3}{2}^-$, and $N(1700)\frac{3}{2}^-$ we have used
two sets of mixing angles: obtained from the description of mass
in Ref. \cite{Isgur2} and 
found in the present work
from the $\pi N$ widths of the resonances.
The calculated amplitudes for the electroexitation of the states $^4 8_{1/2}$
and $^4 8_{3/2}$ turned out  significantly smaller than the
amplitudes for the states $^2 8_{1/2}$
and $^2 8_{3/2}$. As a result, the mixings do not affect practically
the electroexcitation amplitudes for the  $N(1535)\frac{1}{2}^-$ and
$N(1520)\frac{3}{2}^-$, but
play a significant role for the $N(1650)\frac{1}{2}^-$
and $N(1700)\frac{3}{2}^-$.
  
The approximation
of the single quark transition model 
\cite{Hey_Weyers,Babcock_Rosner,Cottingham,SQTM}
leads to selection rules, which for
the resonance $N(1675)\frac{5}{2}^-$ result
in the suppression of the amplitudes
$A_{1/2}(Q^2)$ and $A_{3/2}(Q^2)$ on the proton.
According to our results,
relativistic effects violate this suppression weakly,
and we expect that experimental values of these amplitudes
will be determined mostly by the meson-baryon contributions.
In contrast with proton, 
the predicted 
electroexcitation amplitudes on the neutron
for the $N(1675)\frac{5}{2}^-$ are large.

For the resonances $N(1520)\frac{3}{2}^-$, $N(1535)\frac{1}{2}^-$, and
$N(1675)\frac{5}{2}^-$, experimental data on electroexcitation amplitudes
on the proton are available in wide range of $Q^2$. This allowed us to
present the expected meson-baryon contributions to these
amplitudes at $Q^2<2-3~$GeV$^2$ inferred from the difference
of the LF RQM predictions and the data.
The correspondence between these contributions and the meson cloud
contributions obtained within the EBAC/Argonne-Osaka
coupled-channels analyses
\cite{EBAC,EBAC1,EBAC2,INT,INT1} is discussed in
Sec. \ref{results3}.

{\bf Acknowledgments}.
This work was supported by
the U.S. Department of Energy, Office of Science,
Office of Nuclear Physics, under Contract
No. DE-AC05-06OR23177, and the National Science Foundation, State
Committee of Science of the Republic of Armenia, Grant No. 15T-1C223.

\begin{thebibliography}{999}



\bibitem{CLAS2009} I. G. Aznauryan, et al., 
CLAS collaboration, Phys. Rev. C{\bf 80}, 055203 (2009).

\bibitem{Capstick1995} S. Capstick and B. D. Keister, Phys. Rev. D
{\bf 51}, 3598 (1995).

\bibitem{Aznauryan2007} I. G. Aznauryan,
Phys. Rev. C {\bf 76}, 025212 (2007).

\bibitem{Stave} S. Stave et al.,
Eur. Phys. J. A {\bf 30}, 471 (2006);
Phys. Rev. C {\bf 78}, 025209 (2008).

\bibitem{Sparveris} N.F. Sparveris et al.,
Phys. Rev. Lett. {\bf 94}, 022003 (2005);
Phys. Lett. B {\bf 651}, 102 (2007).

\bibitem{Mertz} C. Mertz et al.,
Phys. Rev. Lett. {\bf 86}, 2963 (2001).

\bibitem{Kunz} C. Kunz et al.,
Phys. Lett. B {\bf 564}, 21 (2003).

\bibitem{Frolov} V.V. Frolov et al.,
Phys. Rev. Lett. {\bf 82}, 45 (1999).

\bibitem{Vilano} A.N. Villano et al.,
Phys. Rev. C {\bf 80}, 035203 (2009).

\bibitem{KELLY} J. J. Kelly et al.,
Phys. Rev. Lett. {\bf 95}, 102001 (2005);
Phys. Rev. C {\bf 75}, 025201 (2007).

\bibitem{Sato2001} T. Sato and T.-S. H. Lee,
Phys. Rev. C {\bf 63}, 055201 (2001).

\bibitem{Lee2004} V. D. Burkert and T.-S. H. Lee,
Int. J. Mod. Phys. E {\bf 13}, 1035 (2004).

\bibitem{EBAC} B. Juli\'a-D\'iaz, T.-S. H. Lee, A. Matsuyama,
T. Sato, and L. C. Smith,
Phys. Rev. C {\bf 77}, 045205 (2008).

\bibitem{Riska2004} B. Juli\'a-D\'iaz, D. O. Riska, and F. Coester,
Phys. Rev. C {\bf 69}, 035212 (2004).

\bibitem{Aznauryan2015} I. G. Aznauryan and V. D. Burkert,
Phys. Rev. C {\bf 92}, 035211 (2015).

\bibitem{Aznauryan2016} I. G. Aznauryan and V. D. Burkert,
arXiv:1603.06692,2015.

\bibitem{CLAS2015} K. Park, et al., 
CLAS collaboration, Phys. Rev. C{\bf 91}, 045203 (2015).

\bibitem{Aznauryan2015_1} I. G. Aznauryan and V. D. Burkert,
Phys. Rev. C {\bf 92}, 015203 (2015).

\bibitem{Thompson} R. Thompson et al.,
CLAS Collaboration,
Phys. Rev. Lett. {\bf 86}, 1702  (2001).

\bibitem{Denizli} H. Denizli et al., CLAS Collaboration,
Phys. Rev. C {\bf 76}, 015204 (2007).

\bibitem{Armstrong} C.S. Armstrong et al.,
Phys. Rev. D {\bf 60}, 052004 (2009).

\bibitem{Dalton} M.M. Dalton et al.,
Phys. Rev. C {\bf 80}, 015205 (2009).

\bibitem{Mokeev1} V.I. Mokeev and I.G. Aznauryan,
Int. J. of Modern Phys., Conf. Series, {\bf 26}, 1460080 (2014).

\bibitem{Mokeev2} V.I. Mokeev et al.,
arXiv:1509.054650[nucl-ex].

\bibitem{Mokeev3} V.I. Mokeev et al.,
CLAS Collaboration, Phys. Rev. C {\bf 86}, 035203 (2012).

\bibitem{CLAS2008} K. Park et al., CLAS Collaboration,
Phys. Rev. C $\bf{77}$, 015208 (2008).

\bibitem{Aznauryan2012} I. G. Aznauryan and V. D. Burkert,
Phys. Rev. C {\bf 85}, 055202 (2012).

\bibitem{Drell} S. D. Drell and T. M. Yan, Phys. Rev. Lett.
{\bf 24}, 181 (1970).

\bibitem{Terentiev} V. B. Berestetskii and M. V. Terent'ev,
Sov. J. Nucl. Phys., {\bf 24}, 1044 (1976);
{\bf 25}, 347 (1977).

\bibitem{Brodsky} S. J. Brodsky and S. D. Drell, Phys. Rev. D
{\bf 22}, 2236 (1980).

\bibitem{Terentiev1} L. A. Kondratyuk and M. V. Terent'ev,
Yad. Fiz., {\bf 31}, 1087 (1980).

\bibitem{Aznauryan1982} I. G. Aznauryan, A. S. Bagdasaryan,
and N. L. Ter-Isaakyan,
Phys. Lett. B {\bf 112}, 393 (1982);
Yad. Fiz. {\bf 36}, 1278 (1982).

\bibitem{Keister} B. D. Keister and W. N. Polizou,
Adv. Nucl. Phys. {\bf 20}, 225 (1991).

\bibitem{Melosh} H. J. Melosh, Phys. Rev. D
{\bf 9}, 1095 (1974).

\bibitem{Isgur1} R. Koniuk and N. Isgur, Phys. Rev. D
{\bf 21}, 1868 (1980).

\bibitem{Aznauryan_review} I. G. Aznauryan, V. D. Burkert,
Prog. Part. Nucl. Phys. {\bf 67}, 1 (2012), arXiv:1109.1720, 2011.

\bibitem{Devenish} R.C.E. Devenish, T.S. Eisenschitz,
and J.G. K\"orner,
Phys. Rev. D {\bf 14}, 3063 (1976).

\bibitem{Isgur2} N. Isgur and G. Karl, Phys. Rev. D
{\bf 18}, 4187 (1978).

\bibitem{Hey} A. J. G. Hey, R. J. Litchfield, and R. J. Cashmore, 
Nucl. Phys. B
{\bf 95}, 516 (1975).

\bibitem{Aznauryan1985} I. G. Aznauryan and A. S. Bagdasaryan,
Yad. Fiz. {\bf 41}, 249 (1985).

\bibitem{PCAC} F. J. Gilman, M. Kugler, and S. Meshkov,
Phys. Rev. D {\bf 9}, 715 (1974).

\bibitem{Hey_Weyers}
A.J.G. Hey and J. Weyers,
Phys. Lett. B {\bf 48}, 69 (1974).

\bibitem{RPP} C. Patrignani et al. [Particle Data Group],
Chinese Physics C {\bf 40}, 100001 (2016).

\bibitem{Babcock_Rosner}
J. Babcock and J.L. Rosner,
Ann. Phys. (N.Y.)  {\bf 96}, 191  (1976) .

\bibitem{Cottingham}
W.N. Cottingham and I.H. Dunbar,
Z. Phys.  C {\bf 2}, 41 (1979).

\bibitem{SQTM}
V.D. Burkert et al.,
Phys. Rev.  C {\bf 67}, 035204 (2003).

\bibitem{Giannini} E. Santopinto and M. M. Giannini,
Phys. Rev. C {\bf 86}, 065202 (2012).

\bibitem{EBAC1} B. Juli\'a-D\'iaz, H. Kamano, T.-S. H. Lee, A. Matsuyama,
T. Sato, and N. Suzuki,
Phys. Rev. C {\bf 80}, 025207 (2009).

\bibitem{EBAC2} H. Kamano, S. X. Nakamura,  T.-S. H. Lee, 
and T. Sato,
Phys. Rev. C {\bf 94}, 015201 (2016).

\bibitem{INT} H. Kamano, talk presented on INT workshop
"Spectrum and Structure of Excited Nucleons 
from Exclusive Electroproduction", Seattle, USA, November, 2016.

\bibitem{INT1} T. Sato, talk presented on INT workshop
"Spectrum and Structure of Excited Nucleons 
from Exclusive Electroproduction", Seattle, USA, November, 2016.

\end{thebibliography}
\end{document}